# Remote Collaboration Fuses Fewer Breakthrough Ideas

**Authors:** Yiling Lin[1], Carl Benedikt Frey[2,3], Lingfei Wu[1,3]

**Affiliations:**

[1] School of Computing and Information, The University of Pittsburgh, 135 N Bellefield Ave, Pittsburgh, PA 15213

[2] Oxford Internet Institute & Oxford Martin School, University of Oxford, 1 St Giles', Oxford OX1 3JS, United Kingdom

[3] These authors jointly supervised this work: Lingfei Wu, Carl Benedikt Frey. e-mail: liw105@pitt.edu; carl.frey@oii.ox.ac.uk

## Abstract
Theories of innovation emphasize the role of social networks and teams as facilitators of breakthrough discoveries[1–4]. Around the world, scientists and inventors today are more plentiful and interconnected than ever before[4]. But while there are more people making discoveries, and more ideas that can be reconfigured in novel ways, research suggests that new ideas are getting harder to find[5,6]—contradicting recombinant growth theory[7,8]. In this paper, we shed new light on this apparent puzzle. Analyzing 20 million research articles and 4 million patent applications across the globe over the past half-century, we begin by documenting the rise of remote collaboration across cities, underlining the growing interconnectedness of scientists and inventors globally. We further show that across all fields, periods, and team sizes, researchers in these remote teams are consistently less likely to make breakthrough discoveries relative to their onsite counterparts. Creating a dataset that allows us to explore the division of labor in knowledge production within teams and across space, we find that among distributed team members, collaboration centers on late-stage, technical tasks involving more codified knowledge. Yet they are less likely to join forces in conceptual tasks—such as conceiving new ideas and designing research—when knowledge is tacit[9]. We conclude that despite striking improvements in digital technology in recent years, remote teams are less likely to integrate the knowledge of their members to produce new, disruptive ideas.

## Introduction

The past half-century has seen a dramatic increase in the scale and complexity of scientific research[4], to which researchers have responded by lengthening their education and training[10], specializing more narrowly[11], and working in teams[2,4,11]. The latter has been aided by recent advances in remote work technology, allowing researchers to form distributed teams to take advantage of complementary yet geographically dispersed knowledge and expertise[12–16]. A widely held view is that by permitting more specialization and better matching, the rise of remote collaboration promises larger "collective brains"[3] and accelerated innovation[7]. Indeed, seen through the lens of recombinant growth theory[7], a larger number of possible collaborations increases the number of possibilities for new discoveries. Yet, in contradiction to this promise, recent work has shown that "ideas are getting harder to find"[5,6].

One possible explanation for this apparent puzzle is that while remote collaboration among specialized researchers permits more novel combinations of knowledge, it also makes it harder for teams to integrate the pieces[17]. In the early stages of a project, when an idea is hard to articulate and knowledge is tacit, collaboration at a distance is particularly challenging[18]. But when an idea crystallizes and knowledge becomes more codified, the comparative advantage of onsite teams is gradually diminished. It follows that scientists in onsite teams are better placed to fuse knowledge and conceive the next breakthrough idea[12,19,20], while they tend to coordinate technical work and develop established ideas when switching to remote[13,21].

In the pages that follow, we show how the roles of team members change as scientists and inventors switch from onsite to remote collaboration. Analyzing 20 million research articles between 1960 and 2020 and 4 million patent applications between 1976 and 2020 across the globe, we confirm that remote teams develop and onsite teams disrupt both in science and technology. Inspired by a recent study linking disruptive innovation to team structure[22], we examine author contribution disclosures and find that despite striking advances in remote work technology, collaboration at a distance still centers on late-stage, technical project tasks rather than conceptual tasks. The tendency of remote teams to execute and not conceptualize is robust to controlling for a host of potential confounders, and seemingly associated with the continued importance of face-to-face interactions. We conclude by showing that established and emerging researchers are much less likely to jointly conceive new ideas when working remotely, reducing the exposure of new talent to disruptive discovery.

Our article makes three key contributions to the existing literature. First and foremost, we shed new light on the deceleration of innovation, despite the rising number of possibilities for discovery and increased research efforts[5]. Shifting the research focus from the performance of individual scientists[6,10,23,24] to their team roles, we show that while remote collaboration involves more people in science and technology, it does not necessarily engage them in the core task of conceiving research. In other words, the creative potential of many researchers, especially emerging scholars, has not been fully realized. Second, although large teams have long been emphasized as a way of mobilizing greater collective knowledge to push the frontiers of science[4,8], recent research shows that small teams and solo researchers are more likely to disrupt both in science and technology[25]. We add to this literature by analyzing interactions within teams and their importance for fusing breakthrough ideas. Finally, while recent research has documented that remote work can increase productivity in routine activities such as in call centers[26], another set of studies shows that it hampers creative activities[20,27,28]. Reconciling these findings, we show how the comparative advantage of remote work shifts as a project progresses. While onsite teams evolve early-stage ideas, remote teams extend established knowledge as it becomes more codified. Taken together, our results point to the critical role that in-person interaction plays in fusing disruptive discoveries and training the next generation of talent in science and technology, even in the age of remote work.

**Research Design**
To compare the innovative performance of onsite teams (with all team members in the same city) and remote teams (with team members spread across two or more cities), we start by creating and analyzing two large datasets representing the full spectrum of science and technology fields, including **(1) scientific research teams** responsible for 20,134,803 papers published by 22,566,650 scientists across 3,562 cities between 1960 and 2020. The name-disambiguated authors and their respective institutions with latitude and longitude values were obtained from the archived version of Microsoft Academic Graph (MAG) and verified in two ways: by two human coders who manually checked a random sample of the data, and by comparing our sample against self-reported records in Open Researcher and Contributor ID (ORCID); **(2) patenting teams** responsible for 4,060,564 patents filed by 2,732,326 inventors across 87,937 cities between 1976 and 2020. The name-disambiguated inventors and their addresses with latitude and longitude values were obtained from PatentsView, an online data platform of the U.S. Patent and Trademark Office (USPTO), and verified by two human coders (see Methods B). These two datasets cover teams of different fields, periods, and team sizes, which allows us to examine the robustness of the relationship between collaboration distance and inventive outcomes when these variables are accounted for. However, without information on what collaborators actually do within teams, it is hard to explain any observed correlation. To overcome this data limitation, we extend our analysis of scientific research teams by including **(3) self-reported author contributions.** Doing so, we collect 89,575 author contribution disclosures published between 2003 and 2020 from the websites of *Nature*, *Science*, *PNAS*, and *PLOS ONE*, and map them to the name-disambiguated scientists in our data. This allows us to provide the first quantitative evidence of how roles change when the same scientist switches from onsite to remote collaboration. We also probe the robustness of our key results in three ways. First, we trace the roles of the same scientists when they work either remotely or onsite. Second, we zoom in on teams that collaborate repeatedly to investigate how roles change when members split geographically as an event study. Third, using machine learning techniques that infer team roles for papers where author contributions are not explicit, thereby increasing our sample size, we check that differences in team roles between onsite and remote teams hold more broadly. We turn to describe these robustness tests in greater detail in Methods J.

For each paper or patent, we calculate a newly proposed yet extensively verified measure, "Disruption" or *D*-score, which assesses to which extent an idea disrupts the state of science or technology[6,25] (see Methods C). Distinguishing between disruptive discoveries and developing ones is crucial, since breakthroughs open up new avenues for progress, while incremental developing projects eventually run into diminishing returns[6]. The intuition of the *D*-score is straightforward: if subsequent work that cites a product also cites its references, the focal product can be seen as building on that prior knowledge. If the converse is true—future works cite a paper or patent but ignore its acknowledged forebears—they recognize that output as disruptive by eclipsing the old ones referenced. *D*-score varies from -1 (developing) to 1 (disruptive) since it is calculated as the difference between the probabilities of observing these two types of subsequent citation patterns[25]. Thus, *D*-score allows us to uncover the distinct roles that research teams play in unfolding the advance of science and technology. For example, the 1953 DNA paper by Watson and Crick[29] is among the most disruptive works ($D = 0.96$, top 1%), whereas the 2001 human genome paper[30] by Watson and others is highly developing ($D = -0.017$, bottom 6%). For robustness, building on the intuition that radical innovation is typically accompanied by new terminology, we also complement our *D*-score measure with a variable identifying papers that proposed new scientific concepts (e.g., "time-evolving block decimation")[31] and patents introducing new technology codes (e.g., "Web crawling techniques for indexing")[32]. We further outline the overall research design and our empirical strategies in Methods A.

**Remote Teams Produce Fewer Breakthroughs**

Over the past half-century, research teams have expanded geographically across all sciences and technology fields (Fig. 1a-d). The average distance between team members has increased from 100km to nearly 1,000km in papers and from 250km to 750km in patents. In tandem, the fraction of extremely long-distance collaborations over 2500km, corresponding to the width of the south Atlantic from Brazil to Liberia, increased substantially from 2% to 15% for papers, and from 3% to 9% for patents (Fig. 2a-c). However, the contribution of remote teams to breakthrough innovation has been far less impressive. Across papers and patents, the probability of disruption P($D$>0) falls from 28% to 22% for papers (p-value < 0.001 for two-side Student's t-test), and 67% to 55% for patents (p-value < 0.001), as collaboration distance increases from 0km to more than 600km—approximately the distance between Paris and Frankfurt (Fig. 3a). In relative terms, the remote work penalty is around 3% ~ 4%, with P($D$>0) declining from 20.4% to 19.5% for papers (p-value < 0.001 for two-side Student's t-test), and 58.2% to 56.5% for patents (p-value < 0.001), when we add our full set of controls, including fields, periods, team sizes, average career age, knowledge diversity, and tie strength, as well as author fixed effects (Extended Data Fig. 1, Extended Data Table. 1-2). This pattern is also robust against alternative measures of collaboration distance (Extended Data Fig. 2)[33] and breakthrough discoveries (Extended Data Fig. 3). Inspired by previous studies on coordination challenges originating both from fewer in-person interactions due to spatial separation and more working schedule conflicts across time zones[27,34], we disentangle these effects and observe a significant decline in the probability of disruption between local teams and remote teams across times zones as well as *within* a time zone (Fig. 3b-c). Overall, our findings consistently point to the continued value of geographic proximity for disruptive innovation.

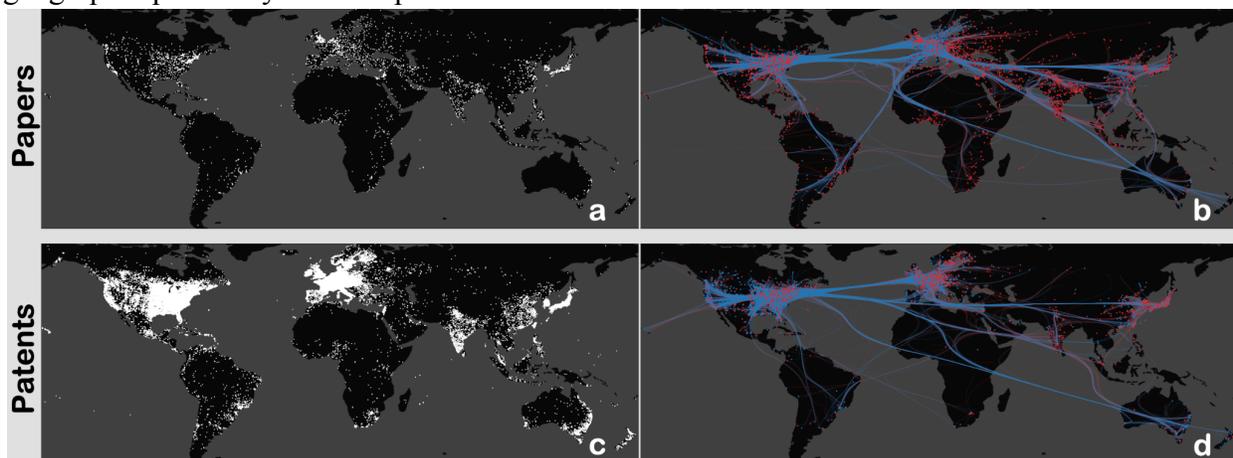

**Fig. 1: Mapping the global collaboration of scientists and inventors.** Our data includes 20,134,803 papers published by scientists across 3,562 cities between 1960 and 2020, as well as 4,060,564 patents filed by 2,732,326 inventors across 87,937 cities between 1976 and 2020. We visualize the geographical distribution of these scientists (**a**) and inventors (**c**). Each dot is a city in our dataset. Note that while there are nearly an order of magnitude fewer patents than papers, there are still over an order of magnitude more patenting cities than paper-producing cities. The greater geographical span of patenting reflects that industry is more dispersed than academia. Building upon (a) and (c), we display where disruptive ($D$>0) papers (**b**) and patents (**d**) are produced. A dot represents all onsite teams based in that city, and an edge between two cities represents all remote teams with members in both cities[35]. The colors of dots and edges indicate whether disruptive work is observed at a higher (red) or lower (blue) probability relative to the population baseline. We analyze dots and edges that contain five or more teams to effectively calculate the probability of observing disruptive work, and find that onsite teams are more disruptive: 76% of cities in science (representing 58% onsite teams) and 48% of cities in patents (representing 76% onsite teams) are red. We also note that remote teams tend to be more developing: 71% of city pairs in science and 63% of city pairs in patents are blue.

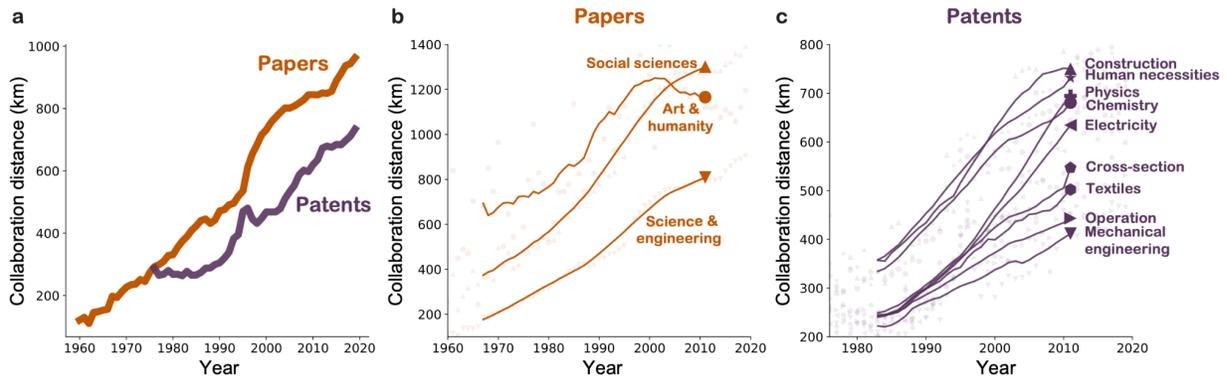

**Fig. 2: The growth of collaboration distance across all sciences and technologies.** We analyzed the geographical distribution of scientists and inventors underlying 20,134,803 research papers published between 1960 and 2020, as well as 4,060,564 patent applications filed between 1976 and 2020. The average distance between co-authors (collaboration distance) has increased dramatically from below 100km to nearly 1,000km for papers and from 250km to nearly 750km for patents during the investigated period (**a**). This increase in collaboration distance holds across all fields for papers (**b**) and technology domains for patents (**c**). In (b) and (c), we display raw data (points) and also the moving average using a long, sixteen-year window (curves). The trends remain the same if alternative window sizes (e.g., two, four, or eight years) are used.

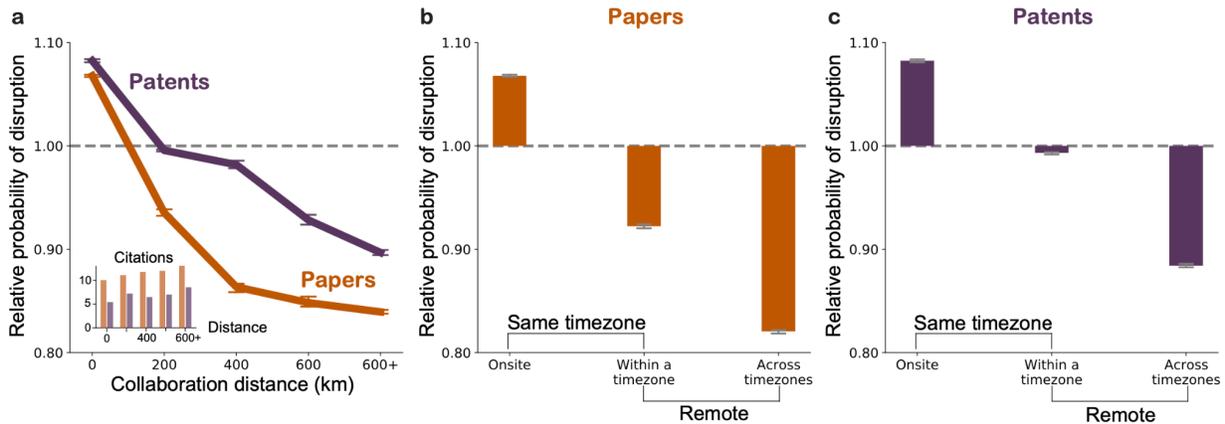

**Fig. 3: Remote teams produce fewer breakthrough innovations.** We analyzed the same datasets as in Fig. 2. We find that the probability of producing disruptive work (D>0) declined from 28% to 22% for papers (p-value < 0.001, n = 14,485,326) and 67% to 55% for patents (p-value < 0.001, n = 3,411,366), respectively, as collaboration distance increased from 0km to over 600km (**a**). In addition, the inset of (a) shows that the average citation impact within 5 years after publication increases with collaboration distance for both papers and patents, confirming that our disruption measure is distinct from citations. We further distinguish between three groups of papers (**b**) and patents (**c**), including local teams as the first group, remote teams within a time zone as the second group, and remote teams across time zones as the third group. We observe a substantial decline in the probability of disruption when moving from local to remote within a timezone (from 28% to 24% for papers and from 67% to 61% for patents), as well as when moving from remote within a timezone to across timezones (from 24% to 22% for papers and from 61% to 55% for patents), with p-values < 0.001 in both comparisons. To facilitate the comparison between papers and patents, we display the relative probability of disruption, which is calculated as the ratio of disruption probability for the group of a given collaboration distance to the disruption probability of the entire population (gray dotted lines). In all panels, the error bars indicate a 95% bootstrap confidence interval centered at the mean. All statistical tests use a two-sided Student's t-test.

**Onsite Ideation, Remote Execution**

With this in mind, we next turn to examine the core hypothesis of this paper: that although remote collaboration permits more novel combinations of knowledge, it also makes it harder for teams to integrate the pieces. Indeed, if maintaining frequent, in-person communication is challenging when team members are spread across cities, and some activities rely more on in-person interaction than others, even the same scientists should change team roles when they switch from onsite to remote collaboration. To test precisely this, we analyze the roles of scientists in teams across four functional research activities, including "conceiving research," "performing experiments," "analyzing data," and "writing the paper." We note that the probability of the same scientist contributing to "conceiving research" declines most dramatically (from 63% to 51%, p-value < 0.001) relative to all other activities (Fig. 4a). On average, scientists in remote teams are less likely to engage in conceptual tasks than their peers in onsite teams (48% vs. 42%, p-value < 0.001), including "conceiving research" or "writing the paper," and correspondingly, more likely to contribute to technical tasks, such as "performing experiments" and "analyzing data" (Fig. 4a inset). These patterns hold when we control for potential confounders, including research fields, periods, and team sizes (see Extended Data Table. 3 and Methods J). In addition, when we switch our focus from the role of individual scientists to their interactions within the team, we find the same pattern: the relative probability of two authors joining forces in conceiving research declines from 34% to 28% (p-value < 0.001) when they switch from onsite to remote collaboration (Extended Data Table. 4).

Our findings have implications for the future of scientific and technological progress. Building on the result that onsite teams involve more talent in conceiving research, we turn to explore how this affects the next generation of researchers, distinguishing between team members by their citation impacts. Doing so, we find that among onsite teams, the probability of two authors joining forces in conceiving research (34%) barely changes with the difference in their citation impacts. However, in remote teams, this probability decreases dramatically from 33%, when two authors have the same level of citations, to 23%, when one has four orders of magnitude more citations than the other (Fig. 4b). The least and most impactful authors, in other words, are much less likely to jointly conceive new ideas in remote teams than in onsite teams (Fig. 4b inset). This striking pattern, whereby onsite teams engage less established researchers in conceptual work, whereas remote teams merely assign them technical tasks, means that in the latter case, the opportunities for idea generation do not trickle down the hierarchy of citation impacts from established scholars to emerging ones. We conclude that onsite teams are particularly important as they serve as an escalator for new talent to co-lead in conceptualizing the next breakthrough.

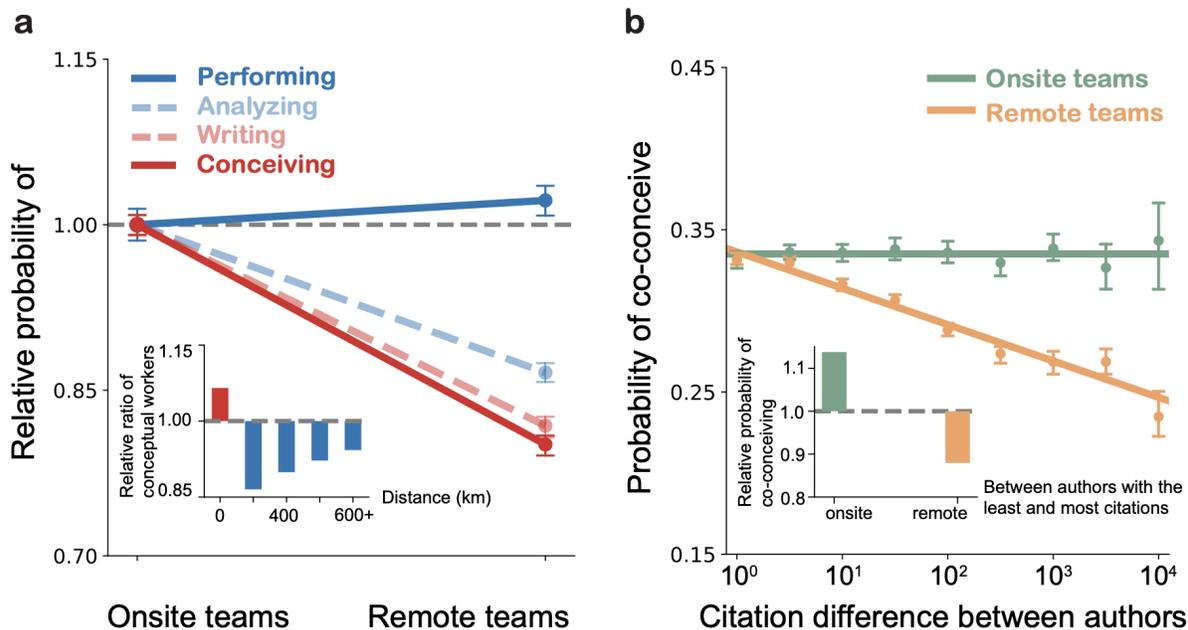

**Fig. 4: Onsite teams conceive, remote teams perform.** (**a**) We analyzed the contribution disclosures of 89,575 papers from *Nature*, *Science*, *PNAS*, and *PLOS ONE* (2003-2020) and identified four research activities. In our sample, the probability of conceiving, writing, and analyzing decreased by 12% (from 63% to 51%), 11% (from 60% to 49%), and 8% (from 58% to 50%), respectively (with p-values < 0.001, n = 21,373), when scientists switched from onsite to remote. Conversely, the probability of performing increased by 1% (from 42% to 43%,; p-value < 0.05, n = 21,373). To ease comparison, we display relative probabilities, or the probability ratio between what the same scientist does in onsite and remote teams. For robustness, we use a larger sample of 16,397,750 papers based on machine learning models (see Methods J) to predict the distinct roles of conceptual workers (engaged in conceiving and writing) and technical workers (engaged in performing and analyzing). Again, we find that scientists are more likely to conceive in onsite teams and less likely to do technical work (inset). (**b**) Based on the contribution statements, we find no relationship between two co-authors joining forces in conceiving research and their citation difference (calculated when their paper was published) among onsite teams (OLS regression indicated by the green line, coefficient = 0.0002, p-value > 0.05, n = 155,842), which might partly reflect more student-advisor relationships. In contrast, among remote teams, the co-conceive probability declines, from 33% to 23%, when the citation difference increases from zero to four orders of magnitude (OLS regression indicated by the orange line, coefficient = -0.022, p-value < 0.001, n = 296,861). When we only consider the co-authors with the least and most citations, we observe a similar decline (inset). In all figures, the 95% bootstrap confidence intervals are shown as error bars centered at the mean. All statistical tests use a two-sided Student's t-test.

## Discussion

In this paper, we shed new light on one of the great puzzles of our time: why the connectivity brought by the Internet has not led to the upsurge in innovation that recombinant theory predicts. Our key finding is that while remote collaboration permits more novel combinations of knowledge in principle, it also makes it harder for teams to integrate the pieces.

At a time when scientific talent is increasingly moving across the globe[36], and workplaces are rethinking their remote work policies in the aftermath of the Covid-19 pandemic, our results have important implications for both managers and policymakers. As we have shown, colocation still plays a key role in the fusion of radical ideas, suggesting that the post-pandemic shift towards remote work will likely favor incremental innovation at the expense of disruptive discoveries. From a managerial point of view, projects aiming at disruptive innovation are best allocated to onsite teams, while projects focusing on incremental improvements can be assigned to their distributed counterparts. Our study also underlines an important trade-off that policymakers face: while remote collaboration might allow for the effective exploitation of existing ideas in the short run, it might also curtail the kind of innovation breakthroughs that drive progress and productivity over the long run. Therefore, for policymakers interested in reviving productivity growth and innovation, physical infrastructure investment to reduce travel costs and make housing affordable where knowledge industries cluster, should not take the backseat to the construction of digital infrastructure.

# Methods

**A. Research design summary**

We begin by tracing the change in collaboration distance, measured as the average geographical distance between the cities of team members, underlying all analyzed papers and patents over the past half-century (Fig. 1-2). We then chart the probability of a team producing innovative breakthroughs, measured by disruption scores, against their collaboration distance (Fig. 3a). We control for a host of potential confounders, including fields, periods, team sizes, average career age, knowledge diversity, tie strength, as well as author fixed effects (Extended Data Fig. 1, Extended Data Table 1-2), and use alternative measures of collaboration distance and breakthrough innovation (Extended Data Fig. 2-3), for robustness. We also disentangle the effects of time and spatial separations to elucidate the disruption decline that directly corresponds to the increase in geographical distance within the same time zone, which points to fewer in-person interactions as a key hurdle to innovation (Fig. 3b-c). To understand why remote teams are less likely to disrupt more concretely, we further investigate how remote and onsite teams organize research activities differently (Extended Data Fig. 4-5), and find that the same scientists tend to lead conceptual tasks onsite, but deliver technical tasks remotely (Fig. 4a). This role change is robust when fields, periods, and team sizes are accounted for (Extended Data Table 3) and holds broadly when we scale up our sample (Fig. 4a inset, Extended Data Fig. 6). We conclude by exploring how the interaction between team members changes from onsite teams to remote teams and find that new talent is much less likely to co-conceive research with established scholars remotely than onsite (Fig. 4b, Extended Data Table 4).

**B. Identifying onsite and remote teams underlying research articles and patent applications**

**Scientific research teams.** MAG provides name-disambiguated authors (22,566,650) and institutions (22,679) of papers based on verified machine learning models[37]. The latitude and longitude values of these institutions are also provided, such as the "University of Pittsburgh" (latitude=40.4445648, longitude=-79.95328) or "Carnegie Mellon University" (latitude=40.44332, longitude=- 79.94358). However, we also verify the quality of MAG's name disambiguation of scientists. Specifically, we selected a random sample of 50 researchers who published 873 papers. Then, two human coders were employed to examine a random sample of 30 papers. All examined papers are confirmed to be correctly assigned, implying 100% accuracy. We also compared all the 873 papers against self-reported publication records downloaded from Open Researcher and Contributor ID (ORCID) and calculated the average recall across 50 scientists as 84%. Again, we employed two human coders to verify the quality of MAG's name disambiguation on institutions and their geographical coordinates, indicating 99% accuracy on the same dataset. Only one incorrect linkage was identified from 131 author-location pairs across 30 papers; the author was assigned to the right research institution but the wrong local branch.

After these verifications, we map the 22,679 institutions to 3,562 cities using the GeoPy API (https://github.com/geopy/geopy). We calculate the geographic distance between coauthors based on the geographic coordinates of cities instead of institutions, so that we identify team members from the same city as onsite, regardless of city size and the distance between institutions within the city. This way, the collaboration distance between a scientist from the University of Pittsburgh and another from Carnegie Mellon University is 0km, as both institutions are located in Pittsburgh, PA. In contrast, the collaboration distance between a scientist from the University of Pittsburgh and a team member from the Massachusetts Institute of Technology is 916km, representing the geographic distance between the centers of Pittsburgh, PA, and Cambridge, MA. Of the papers studied, 68% of authors are in the same city, while 32% are distributed across cities. Among remote teams, 22% of the sample have a collaboration distance of 0 to 200km, while 11%, 7%, and 60% of the papers have a collaboration distance of 200km to 400km, 400km to 600km, and over 600km, respectively.

**Patenting teams.** PatentsView is a patent data-sharing initiative supported by the Office of the Chief Economist in the United States Patent and Trademark Office (USPTO). It provides name-disambiguated authors of patents and their respective residences. The corresponding latitude and longitude values associated with the reported addresses are also provided by PatentsView. We conducted extensive entity matching work to disambiguate city names (e.g., both Osaka-fu and Osaka are represented as Osaka, Japan) and merge different versions of geographic coordinates under the same city. This way, we obtain 87,937 cities of unique latitude and longitude values. Finally, we verify the quality of PatentsView's name disambiguation of inventors as follows. We selected a random sample of 50 inventors who filed 1,975 patents. Two human coders then examined a random sample of 30 patents from the selected patents. All examined USPTO patents are confirmed to be correctly assigned, giving us 100% accuracy. In the same way, we also verify the quality of PatentsView's name disambiguation on residences and their geographical coordinates with 100% accuracy. No incorrect linkages were identified among the 98 inventor-location pairs across 30 patents.

The archived version of MAG we downloaded included 245,253,596 entities, among which 166,274,891 entities have known types, including journal articles (87,285,913), patents (58,972,869), thesis (5,204,930), conference papers (4,803,560), books (4,373,655), book chapters (3,795,548), repositories (1,715,435), and datasets (122,981). We create our paper dataset by selecting the 87 million journal articles, the largest category of scientific papers. We did not combine them with conference papers or theses, as different categories might follow different citation practices that make the calculated disruption score hard to compare. We then select the papers that have two or more authors to focus on teamwork. This leaves us with 58 million papers. We also restrict our sample to papers where all scientists have provided their affiliation information, so we can retrieve author cities and distinguish between onsite and remote teams. This leaves us with 22 million papers. Finally, we keep the papers where each author only provides one affiliation to ensure that the retrieved location information is precise. We are left with 20 million papers.

We also experimented with four different versions of the distance threshold to distinguish between onsite teams (that is, teams with a collaboration distance equaling or below the threshold) and remote teams (with a collaboration distance above the threshold), including 0km, 1km, 5km, and 10km. Specifically, we consistently map scientists to cities before calculating the geographical distance between them. Therefore, these different distance thresholds apply to the distance between cities. We find that the reduced disruption in remote teams is robust across these thresholds. For simplicity, we use the 0km measure, meaning that all team members are in the same location. We use this definition of onsite teams throughout the paper unless specified otherwise. We note that while there are nearly an order of magnitude fewer patents than papers, there are more than an order of magnitude more patenting cities than paper-producing cities. The main reason why the patent dataset contains less teamwork but more cities is that paper authors are highly concentrated in universities, which cluster in large cities or campus towns. In contrast, the greater geographical span of patenting reflects that industry is more dispersed than academia. To ensure the quality of both datasets, we verified that the identified 87,937 patenting cities are all unique addresses—which excludes the possibility that their total number is incorrectly inflated due to repeated records. We also find that the patenting cities span not only the majority (95%) of the large cities and campus towns included in the 3,562 paper-producing cities, but also many other smaller towns that are not included (e.g., in Central Africa). Of the patents examined, 25% of authors are in the same city, while 75% are distributed across cities. Among remote teams, 70% of our sample have a collaboration distance of 0 to 200km, while 7%, 3%, and 20% of the patents have a collaboration distance of 200km to 400km, 400km to 600km, and over 600km, respectively.

### C. Calculating *D*-scores

Subsequent research can reference the primary work in three ways: ($i$) citing only the focal work, ($j$) citing both the focal work and its references, or ($k$) citing only the references. The "Disruption" or *D*-score of a focal paper, denoted as *D*, can be quantified by analyzing the divergence between two categories of subsequent papers:

$$D = p_i - p_j = (n_i - n_j) / (n_i + n_j + n_k)$$

where $p_i$ is the proportion of papers solely referencing the focal paper without including its references, while $p_j$ is the proportion of papers that reference both the focal paper and its associated references. A paper may disrupt earlier research by introducing new ideas that come to be recognized independently from the prior work ($0 < D < 1$), develop existing research by providing supportive evidence or extensions that come to be recognized as incremental additions to prior work ($-1 < D < 0$), or remain neutral, meaning that the disruptive and developmental character of its contribution balances out ($D = 0$).

### D. Quantifying timezone differences underlying inventive teamwork

For 20,273,444 papers and 3,709,940 patents, we map the timezone of each team member based on the latitude and longitude values of their respective cities using PYTZ, a Python API (pytz.sourceforge.net/). We then calculate the hour differences between these time zones. For each team, we calculate the average time zone difference between all pairs of team members as a proxy for the underlying temporal separation.

### E. Identifying the fields of study for research articles and patent applications

**Research articles.** We rely on the scientific taxonomy published by the Microsoft Academic Graph (MAG) team, consisting of a six-level hierarchy. The level-zero labels cover 19 research fields, such as "Mathematics," "Biology," and "Chemistry," while level-one labels cover 292 subfields, and levels two to five labels contain 543,454 unique keywords or phrases. Each MAG paper is linked to one or more labels based on a machine-learning model developed and verified by the MAG team[38]. Within a paper, each label is also associated with a probability value between zero and one that reflects the confidence level of machine prediction. In our analysis, we use the level-zero label of each paper, and if a paper has two or more labels, we select the one with the highest confidence level.

**Patent applications.** The technological taxonomy included in the PatentsView data is the Cooperative Patent Classification (CPC), a four-level classification system. The level-zero has nine sections, including, for example, "Mechanical Engineering; Lighting; Heating; Weapons; Blasting" in Section F, and "Performing Operations; Transporting" in Section B. Under these nine sections, the CPC also provides 128 subsections, 666 groups, and 229,109 subgroups. Each patent has multiple labels that may span across these four levels. For each patent, we first assign all labels to one of the nine section labels at level zero and then select the most popular section label.

**F. Quantifying knowledge diversity of inventive teamwork**
We calculate the interdisciplinarity of team members and use it as a proxy for the diversity of knowledge to which the team has access. Importantly, this allows us to account for team heterogeneity in our regression analysis (Extended Data Table 1-2). To construct this measure for research articles, we first identify the "home discipline" of a scientist from the nineteen top-level MAG field-of-study labels such as "Mathematics," "Biology," and "Chemistry," if they have published three or more papers over half of which are within a single field-of-study. We then distinguish monodisciplinary teams, where all team members are from the same home discipline, from interdisciplinary teams, where team members are from different disciplines. Across the 7,883,633 research teams for which this variable was constructed, we find that the probability of interdisciplinary collaboration is higher in remote teams than in onsite teams (35.6% vs. 28.9%). Leveraging the CPC classification, we apply the same computational method to patent applications and construct disciplinary/interdisciplinary labels for 1,752,307 innovating teams. Again, remote teams outperform onsite teams in interdisciplinary collaboration (19.3% vs. 19.2%). These findings support the view that remote teams are more heterogeneous than onsite teams[1,14,16].

**G. Quantifying tie strength within research and innovation teamwork**
We construct a social network comprising 22,566,650 scientists and 67,226,924 co-authoring relationships using our dataset of research articles. Following a recent study[39], we calculate the strength of the tie between two scientists as the ratio of their common collaborators to their total collaborators. This measure, in other words, defines tie strength as the extent to which two scientists share their collaborators. However, one limitation of this approach is that it may overrate the tie strength between pairs of scientists who only published one or two papers with the same set of co-authors. To address this issue (and for consistency with our analysis of knowledge diversity), we focus on scientists who published three or more papers, over half of which are within a single field-of-study, when calculating the tie strength. We then distinguish weak ties (below the median) from strong ties (above the median). If a team contains one or more weak ties between co-authors, we label this team as a "weak-tie collaboration." Doing so, we find that the probability of weak-tie collaboration is higher in remote teams than in onsite teams (94.4% vs. 82.1%). Leveraging the same computational model, we calculate tie strength on the social network comprising 2,732,326 inventors and 10,476,225 co-authoring relationships using our dataset of patent applications. Again, the probability of weak-tie collaboration is higher for remote teams than onsite teams (76.4% vs 73.1%). These observations support the view that remote teams include more knowledge brokers and weak ties than onsite teams[1,14].

**H. Evaluating the robust, negative relationship between remote teams and disruption**
We run several regressions to evaluate the negative relationship between remote teams and disruption. From our dataset of scientific teams, we selected 7,681,669 scientists who published two or more papers. These scientists have published 13,711,470 papers between 1960 and 2020, which yields 45,078,179 paper-author records. We use this dataset to build stepwise regression models and explore the relationship between remote collaboration (the value equals one if the team members are spread across cities, zero otherwise) and disruption, starting from a model without any control variables or fixed effects. We then add controls for team size, period, average career age, knowledge diversity, tie strength, the field of study, and author-fixed effects. These linear models, inspired by previous studies[40], confirm that remote teams are consistently less disruptive than onsite teams for papers (Extended Data Table 1). From our dataset of patenting teams, we selected 1,253,090 inventors who filed two or more patents. These inventors have filed 2,903,964 patents between 1976 and 2020, which yields 9,031,126 patent-author records. We use this dataset to build stepwise regression models in the same way as mentioned above. Among patents, we confirm a robust, negative relationship between remote teams and disruption when control variables and author-fixed effects are included (Extended Data Table 2). Finally, we show that the lower innovative performance of remote teams is robust against the interaction between the remote collaboration and periods.

**I. Identifying author contributions to scientific papers**
Our author contribution data covers 89,575 contribution disclosures collected from the website of four journals, including *Nature*, *Science*, *PNAS*, and *PLOS ONE*, between 2003 and 2020. Following existing studies[22,41], we identify four functional research activities from contribution statements using Natural Language Processing techniques, including "conceiving research," "writing the paper," "performing experiments," and "analyzing data." We then link authors with their contributions classified into these four categories. We also note that these four categories cohere into two broad roles, including (1) conceptual work leaders, who conceive research and write papers, and (2) technical work supporters, who perform experiments and analyze data.

**J. Evaluating the robust, negative relationship between remote teams and conceiving research**
To investigate how scientists interact differently when the collaboration distance between them increases, we compare the team role of the same scientists when in onsite and remote teams. For 21,373 scientists who published both onsite and remote team papers, their average probability of contributing to "conceiving research" is 63% in onsite teams and 51% in remote teams. We confirm that this decline is statistically significant (p-value < 0.001 for the Student's t-test). For comparison, their probability of "writing the paper" and "analyzing data" decreased by 11% (from 60% to 49%, p-value < 0.001) and 8% (from 58% to 50%, p-value < 0.001), respectively, whereas the probability of "performing experiments" increased by 1% (from 42% to 43%, p-value < 0.05), when switching from onsite to remote collaboration.

We next verify the relationship between this role shift—from leading conceptual tasks to delivering technical tasks—and collaboration distance in three ways. First, among the 21,373 scientists who worked in both onsite and remote teams, our regression analysis confirms the robustness of their role change when fields, periods, and team sizes are accounted for (Extended

Data Table 3). Doing so, we note that the reduced engagement in conceptual tasks in remote teams can not be explained by their larger team size, distinct research fields, or time periods.

Second, to assess the impact of collaboration distance on team roles, we conducted a team-split analysis of the same group of scientists who repeatedly collaborated before and after team members moved. For 15,294 pairs of scientists, across all team sizes, who collaborated in both onsite and remote teams, their probability of jointly contributing to "conceiving research" decreases from 33.5% in onsite teams to 28.3% in remote teams (p-value < 0.001). The reduction in co-conceiving probability is robust when fields, periods, and team sizes are accounted for (Extended Data Table 4). For 2,343 groups of three or more scientists who published in both onsite and remote teams, their probability of contributing to "conceiving research" is 21.6% in onsite teams and 17.8% in remote teams (p-value < 0.01).

Third, we build machine learning models that effectively infer team roles for papers with implicit author contributions to examine if the difference in team roles between onsite and remote teams hold in a much larger sample. Specifically, using our data of author contributions, we train a neural network to infer the two distinct author roles of interest—i.e., leading conceptual tasks and delivering technical tasks—across 16,397,750 papers. These papers are selected from the 20,134,803 papers in our sample, and based on the condition that each selected paper contains variables that are used in the machine learning model. Specifically, we use eight different variables to predict the dichotomy of author roles, including 1) contribution to references, defined as the overlap between references of the focal paper and all references across previously published papers for each author; 2) contribution to topics, defined as the overlap between MAG topic keywords for the focal paper and all keywords across previously published papers for each author; 3) contribution to leading the research, defined as the probability of being the first author(s); 4) contribution to managing correspondence and presentation, defined as the probability of being the corresponding author(s); 5) career age, defined as the number of years from the first publication to the publication of the focal paper for a given author, 6) citation impact, defined as the total number of citations an author has received to all previous publications; 7) topic diversity, defined as the total number of unique MAG topic keywords across previous publications, and finally; 8) publication productivity, defined as the total number of previous papers until the publication of the focal paper. The missing papers did not have these variables for all authors. The machine learning model gives a precision of 0.79 and a recall of 0.793 in predicting author roles. The predicted and empirical values of the fraction of conceptual workers in ground-truth data are highly correlated (Pearson correlation coefficient 0.66, p-value < 0.001). Analyzing these inferred author roles, we find that remote team members are less likely to contribute to conceptual work than their peers in onsite teams (42% vs. 48%) and, correspondingly, more likely to contribute to technical work.

**K. Examining alternative explanations for the reduced disruption of remote teams**
We also consider several alternative explanations for the negative relationship between collaboration distance and idea disruptiveness.

**Team size effect.** Previous work has shown that large teams are less likely to make disruptive discoveries[22,25]. This finding raises the concern that systematic size differences between remote and onsite teams might drive our results, not least if remote teams have grown faster over the investigated period. In response to this concern, we first compare the size of remote and onsite teams over time and confirm that the size of remote teams has grown faster than onsite teams for both papers and patents. Specifically, the average team size increased by 100% (from 2.6 to 5.2) among remote teams but only by 65% among onsite teams (from 2.6 to 4.3) in papers. The same pattern holds for patents: the average team size increased by 40% (from 2.7 to 3.9) among remote teams, but it only increased by 32% among onsite teams (from 2.5 to 3.3). However, we also find that accounting for both team size and periods in our regression models does not alter the negative coefficient of remote teams (Extended Data Table 1-2). These findings lead us to conclude that the difference in average team size or growth rate is unlikely to fully explain the observed differences in inventive output between onsite and remote teams.

**Team composition effect.** Remote teams might also differ from onsite teams in their composition of diverse expertise. Previous research has suggested that remote teams might be more heterogeneous, as geographically distant ties serve as channels for diverse knowledge[1,14,16]. From this perspective, the reduced disruptiveness of remote teams could simply reflect the challenge of integrating more diverse knowledge, regardless of distance. To address this concern, we calculate team member interdisciplinarity and use it as a proxy for diverse knowledge to which the team members have access. We confirm that remote teams are more heterogeneous than onsite teams[1,14,16]. We then include the constructed variable in our regression models and find that the negative impact of remote teams on disruption remains intact (Extended Data Table 1-2). We conclude that differences in team heterogeneity are unlikely to explain the observed difference between onsite and remote teams.

**Age effect.** Previous research has associated the innovation performance of scientists with their age. On the one hand, if acquiring a certain amount of knowledge is a prerequisite for a breakthrough[10], then age and working experience are likely to contribute towards more important discoveries. On the other hand, aging scholars might experience "cognitive entrenchment"[23], and established scholars could become gatekeepers against new ideas[24,42]. In both scenarios, the age differences between remote and onsite teams present a potential cofounder against collaboration distance underlying the reduced disruptiveness of remote teams. Consistent with this reasoning, onsite team members have lower career ages than remote team members on average (9.6 vs. 11.8), possibly reflecting a greater prominence of student-advisor relationships among onsite teams. However, when we include career age in our regression analysis, the negative impact of remote teams on disruption remains unchanged. We conclude that the age structure of remote and onsite teams cannot account for our key findings.

**Selection bias.** Another possibility is that more creative scientists are part of onsite teams rather than remote teams. If this is true, the observed reduced disruptiveness of remote teams originates from differences in individual characteristics rather than from individuals interacting and collaborating in different ways in remote and onsite teams. To that end, we note that the same scientists act differently across team contexts—they are more likely to "conceive research" and "write papers" in onsite teams, and more likely to "perform experiments" and "analyze data" in remote teams, as Fig.4a shows. Second, to further mitigate concerns over selection bias, we run author-fixed effects regressions and confirm that the negative impact of remote teams is still statistically significant (Extended Data Table 1-2), though the magnitude of the coefficient is reduced, possibly because less disruptive scholars end up at more marginal universities, where they benefit more from the opportunities for remote collaboration. We conclude that selection or individual differences cannot fully explain the observed difference between onsite and remote teams.

**Weak tie effect.** Remote teams are likely to include more distant, "weak ties"[43] between team members than onsite teams. However, the impact of these weak ties on innovative performance remains unclear. On the one hand, brokers (i.e., people with diverse and distant contacts) tend to contribute to team innovation because they have access to more diverse knowledge[14,39]. At the same time, these brokers also tend to do worse in gathering the support or interest of their colleagues in delivering innovative ideas[1]. To explore the role of weak ties in team innovation, we quantified tie strength in both publishing and patenting teams and confirmed that remote teams include more weak ties than onsite teams[1,14]. We then include a binary variable of "weak-tie collaboration" in our regression model and confirm that while "weak ties" are associated with more disruptive discoveries, the negative relationship between distance and disruption remains intact (Extended Data Table 1-2). Hence, even though remote teams have access to more diverse knowledge through weak ties, they fail to exchange, fuse, and integrate that knowledge to generate disruptive ideas.

**Acknowledgments**

The authors thank Fengli Xu, James A. Evans, Frank Neffke, Junming Huang, Giorgio Presidente, Hyejin Youn, Yuru Lin, Morgan Frank, and Sarah Bana for helpful comments or discussions. We are grateful for support from Citi, the Dieter Schwarz Foundation (C.B.F), the Richard King Mellon Foundation, the Alfred P. Sloan Foundation, and the National Science Foundation grant SOS:DCI 2239418 (L.W.).

**Author contributions**

Y.L., C.B.F., and L.W. collaboratively conceived and designed the study, contributed to the interpretation of data, and drafted, revised, and edited the manuscript. Y.L. analyzed the data and implemented the models.

**Competing interests**

The authors declare no competing interests.

**Additional information**

**Supplementary information**

**Correspondence and requests for materials**

Correspondence and requests for materials should be addressed to Lingfei Wu and Carl Benedikt Frey.

**Code availability**

All code is available at https://lyl010.github.io/.

**Data availability**

The datasets used in this paper are available at https://lyl010.github.io/ and https://doi.org/10.6084/m9.figshare.21295725.

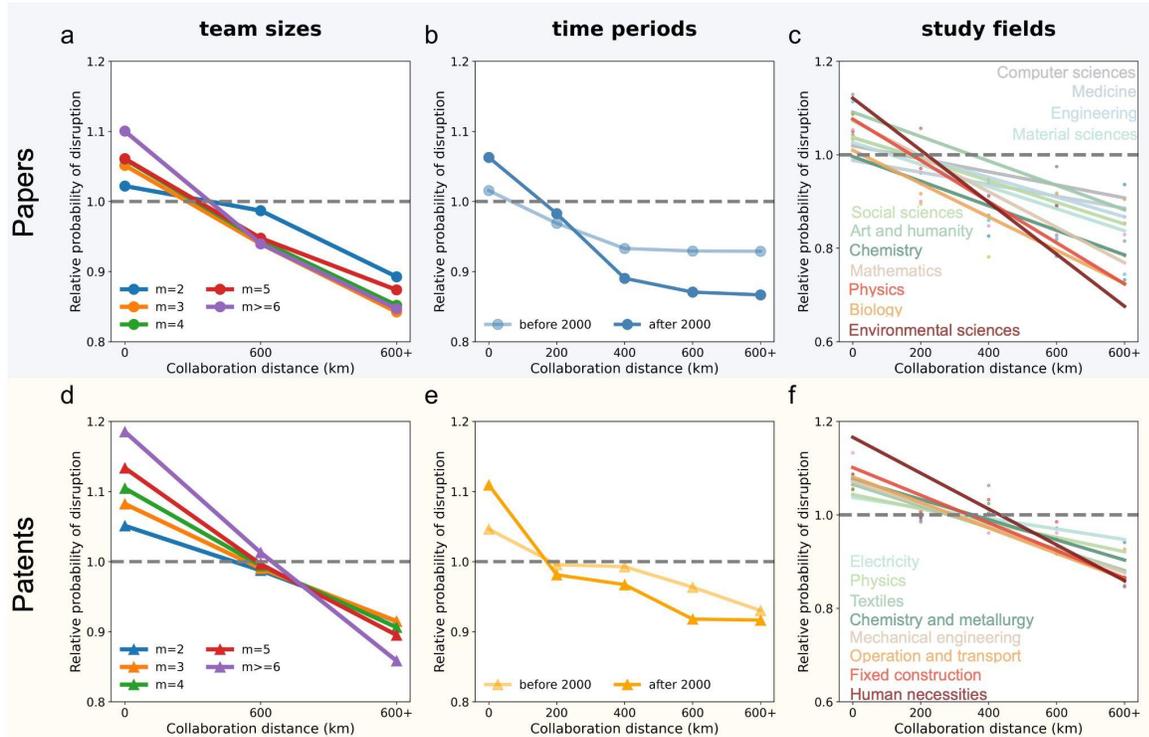

**Extended Data Fig. 1. The robust, negative relationship between collaboration distance and disruption probability against team sizes, periods, and fields.** For 20,134,803 research papers published between 1960 and 2020, and 4,060,564 patent applications filed between 1976 and 2020, we show that the negative association between collaboration distance and the probability of producing disruptive work (D>0) is robust against team sizes (**a, d**), team periods (**b, e**), and fields of study (**c, f**). For each value of the controlled variable (e.g., team size equals two for the blue curve in Panel a), the gray dotted line marks the average disruption probability across all distances. We then display the relative probability of disruption (colored curves), calculated as the ratio of disruption probability for the group of given collaboration distance to the average probability across all distances. The plotted curves have been normalized by dividing their raw values by the group mean, so that the intercepts are not meaningful. We note that Panels b and e appear to show that the remote penalty has strengthened after 2000. However, this effect is confounded by the increase in team size over this time period. See Extended Data Table 1-2 for the effect of distance on innovation when a host of control variables are accounted for.

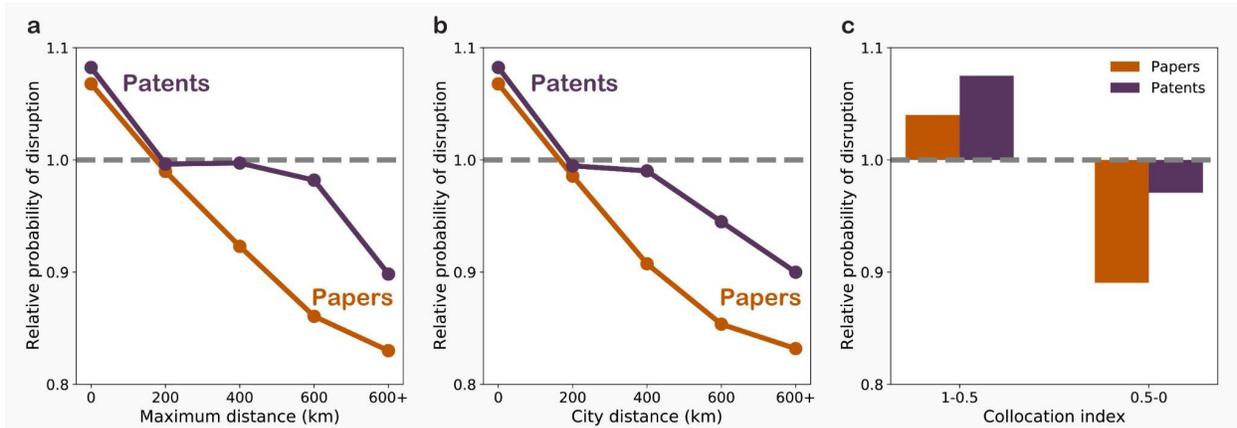

**Extended Data Figure 2. Verifying the lower performance of remote teams using alternative measures of collaboration distance.** For 20,134,803 research papers published between 1960 and 2020, and 4,060,564 patent applications filed between 1976 and 2020, we calculate three different, alternative measures of collaboration distance other than our main specification—the average geographic distance between team members. These include the maximum distance between team members[33] (**a**); the average distance between the unique cities where team members are located (**b**); and a colocation index varying from zero to one, which measures the probability that a randomly selected pair of team members are in the same location[44] (**c**). This colocation index is a continuous variable that complements the binary measures of onsite and remote teams, as it captures boundary cases where some but not all members of a remote team are onsite. In Panel **a-c**, the gray dotted lines mark the average disruption probability for papers and patents across all distances. The colored curves (or bars) mark the relative probability of disruption, calculated as the ratio of disruption probability for the analyzed group to the average probability across all distances.

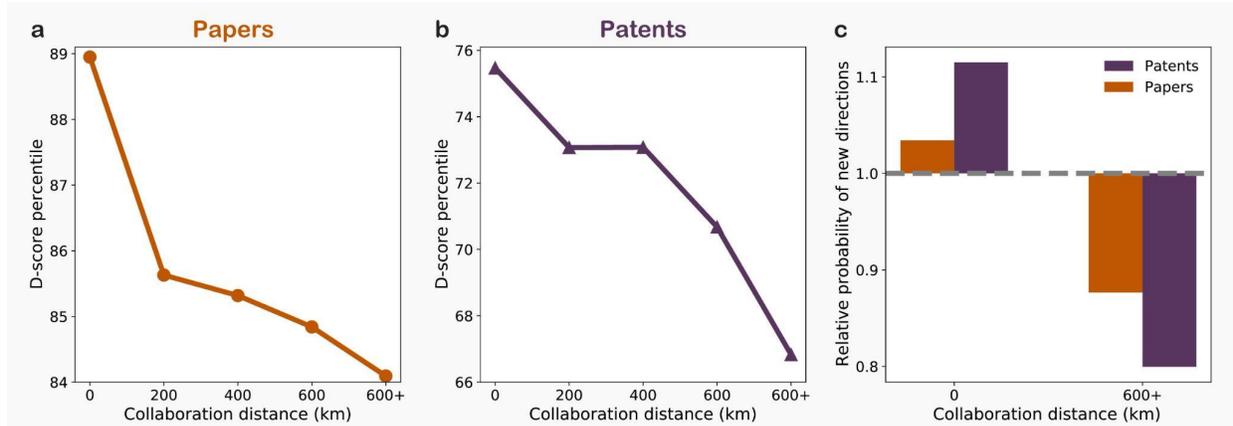

**Extended Data Figure 3. Verifying the lower performance of remote teams using alternative measures of innovation.** For 20,134,803 research papers published between 1960 and 2020, and 4,060,564 patent applications filed between 1976 and 2020, we calculate two alternative measures of innovation and find that remote teams are consistently less likely to disrupt science and technology than onsite teams. The percentile of the average *D*-score falls from 89 to 84 for papers (**a**) and from 76 to 67 for patents (**b**) across the full sample. The probability of proposing new scientific concepts decreases from 0.40% to 0.32% for papers and the probability of introducing new technology codes decreases from 3.33% to 3.22% for patents, when switching from onsite to remote (**c**). The gray dotted line marks the probability of introducing new concepts or code for an average paper (0.37%) or patent (3.24%). The color bars show the relative probability, calculated as the ratio of probability for the analyzed group relative to the population's average probability.

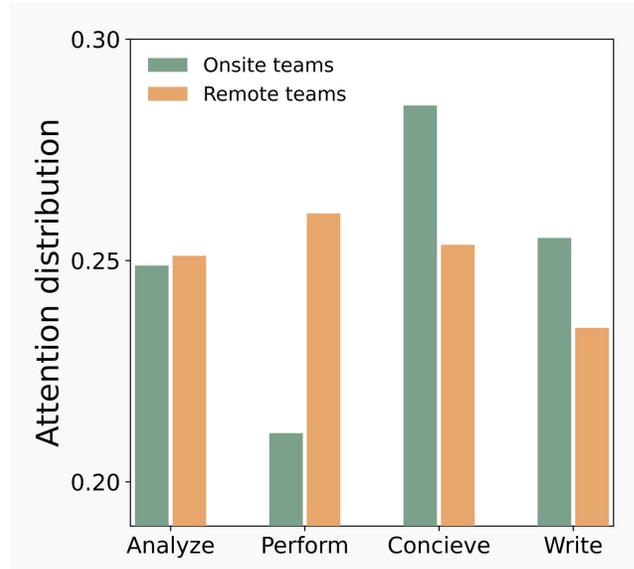

**Extended Data Fig. 4. Onsite teams conceive, and remote teams perform.** We analyzed 89,575 author contribution disclosures underlying papers across four journals, including *Nature*, *Science*, *PNAS*, and *PLOS ONE*, between 2003 and 2020. We associate authors with their contribution to four research activities, including "conceiving research," "performing experiments," "analyzing data," and "writing the paper." For each of the 21,373 scientists who worked in both onsite and remote teams, we track the distribution of their contributions across four activities within each paper and average this distribution within the onsite-team and remote-team papers they published, respectively. This way, we obtained two distributions for each scientist. We then averaged these two distributions across all the 21,373 scientists in our sample. Finally, we displayed the obtained distributions using the Gaussian kernel density estimate. We observe that the key contribution of the same scientist, marked by the peak of the density curves, shifted from "conceiving research" to "performing experiments" when they switched from onsite to remote. These two distributions are significantly different from each other (Chi-squared test statistic = 3188, p-value < 0.001).

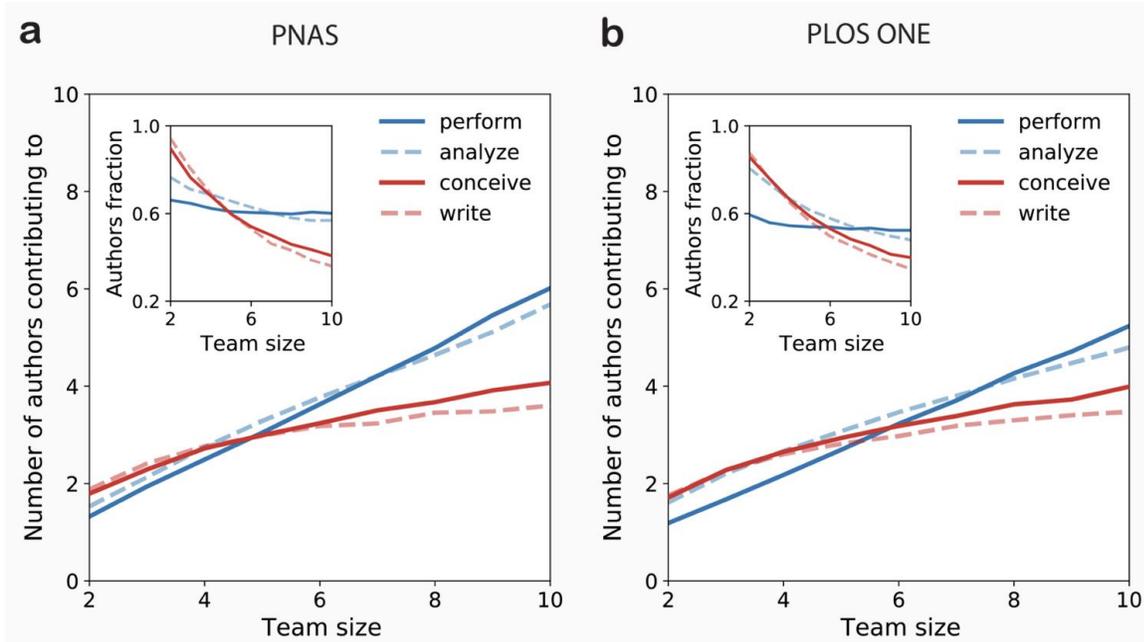

**Extended Data Fig. 5. Different scaling patterns of conceptual and technical activities.** Our author contribution data covers disclosures from *PNAS* (18,354), *Nature* (9,364), *Science* (1,176), and *PLOS ONE* (60,681) between 2003 and 2020. We select *PNAS* (**a**) and *PLOS ONE* (**b**), which have the most observations, and explore the distinct scalability of engagement across research activities. We group four research activities into two broad categories based on their different scalability, including 1) conceptual tasks that contain conceiving and writing, which scale up slowly with team size (red curves), and 2) technical tasks comprising performing and analyzing, which scale up fast with team size (blue curves). As shown in the insets, while the fraction of performing members stabilizes at 0.6 as the team size increases from two to ten, the fraction of conceiving members even decreases from 0.9 to 0.4.

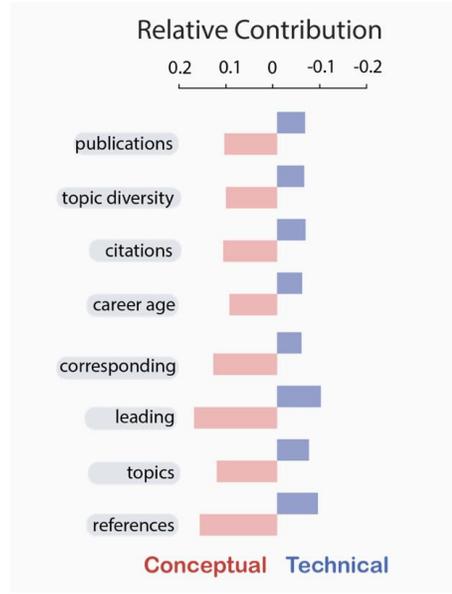

**Extended Data Fig. 6. Inferring conceptual and technical activities.** Using the ground-truth dataset mentioned in Extended Data Fig. 5, we train a neural-network model to infer these two author roles within 16,397,750 papers where author contributions are not explicit. This machine learning model uses eight different variables to predict the dichotomy of author roles, including 1) contribution to references, defined as the overlap between references of the focal paper and all references across previously published papers for each author; 2) contribution to topics, defined as the overlap between MAG topic keywords for the focal paper and all keywords across previously published papers for each author; 3) contribution to leading the research, defined as the probability of being the first author(s); 4) contribution to managing correspondence and presentation, defined as the probability of being the corresponding author(s); 5) career age, defined as the number of years from the first publication to the publication of the focal paper for a given author, 6) citation impact, defined as the total number of citations an author has received to all previous publications; 7) topic diversity, defined as the total number of unique MAG topic keywords across previous publications, and finally; 8) publication productivity, defined as the total number of previous papers until the publication of the focal paper. The machine learning model gives a precision of 0.790 and a recall of 0.793. The predicted and empirical values of the fraction of conceptual workers are highly correlated (Pearson correlation coefficient 0.66, P-value < 0.001). In (c), the eight predictors and their contribution to the prediction are displayed. The figure below is reproduced from our earlier research[22].

|  | D-score > 0 | | | | | | |
| --- | --- | --- | --- | --- | --- | --- | --- |
|  | Model 1 | Model 2 | Model 3 | Model 4 | Model 5 | Model 6 | Model 7 |
| Remote team | -0.0484*** (0.0003) | -0.0380*** (0.0003) | -0.0262*** (0.0003) | -0.0259*** (0.0003) | -0.0146*** (0.0002) | -0.0086*** (0.0003) | -0.0127*** (0.0007) |
| Team size (2~10) |  | -0.0144*** (0.0001) | -0.0105*** (0.0001) | -0.0106*** (0.0001) | -0.0090*** (0.00004) | -0.0149*** (0.0001) | -0.0149*** (0.0001) |
| After 2000 |  |  | -0.1030*** (0.0003) | -0.1029*** (0.0003) | -0.0934*** (0.0002) | -0.0758*** (0.0005) | -0.0773*** (0.0005) |
| Average career age (0~50) |  |  |  |  |  | -0.0008*** (0.00003) | -0.0008*** (0.00003) |
| Interdisciplinary team |  |  |  |  |  | 0.0111*** (0.0004) | 0.0111*** (0.0004) |
| Weak-tie team |  |  |  |  |  | 0.0056*** (0.0004) | 0.0056*** (0.0004) |
| After2000*Remote team |  |  |  |  |  |  | 0.0049*** (0.0007) |
| Remote team + After2000*Remote Team |  |  |  |  |  |  | -0.0078*** (0.0003) |
| Field controls |  |  |  | ✓ | ✓ | ✓ | ✓ |
| Author fixed effects |  |  |  |  | ✓ | ✓ | ✓ |
| Constant | 0.2746*** | 0.3276*** | 0.3841*** | 0.3847*** | 0.3551*** | 0.3315*** | 0.3317*** |
| N. of Observations | 13,711,470 | 13,711,470 | 13,711,470 | 13,711,470 | 45,078,179 | 13,562,214 | 13,562,214 |
| R-squared | 0.0027 | 0.0065 | 0.0167 | 0.0168 | 0.2648 | 0.2933 | 0.2933 |

**Extended Data Table 1. Assessing the robustness of declined disruption with increased collaboration distance in science**. From our dataset of scientific teams, we selected 7,681,669 scientists who published two or more papers. These scientists have published 13,711,470 papers between 1960 and 2020, yielding 45,078,179 paper-author records. We use this dataset to build stepwise regression models and explore the robustness of the relationship between remote collaboration (the value equals one if team members spread across cities, zero otherwise) and disruption, starting from a model without any controls, and then adding team size, time period, average career age, knowledge diversity, tie strength, the field of study, author fixed effects, and finally, an interaction term between time and remote. We note that the remote work penalty—the negative relationship between remote collaboration and disruption—is robust across all specifications. When teams move from 0km to more than 600km collaboration distance, for example, the predicted disruption probability, holding other variables constant, declines from 20.4% to 19.5% (p-value < 0.001 for two-side Student's t-test), or 4.4% in relative terms.

Note: All statistical tests are two-sided t-test and no adjustments were made for multiple comparisons. For Model 5-7, standard errors (in parentheses) are clustered at the author level. * p < 0.05; ** p-value < 0.01; *** p-value < 0.001. We used the REGHDFE package in STATA16[45] to implement the fixed-effects regressions.

|  | D-score > 0 | | | | | | |
| --- | --- | --- | --- | --- | --- | --- | --- |
|  | Model 1 | Model 2 | Model 3 | Model 4 | Model 5 | Model 6 | Model 7 |
| Remote team | -0.0746*** (0.0007) | -0.0706*** (0.0007) | -0.0758*** (0.0007) | -0.0580*** (0.0006) | -0.0050*** (0.0005) | -0.0080*** (0.0009) | -0.0056*** (0.0016) |
| Team size (2~10) |  | -0.0084*** (0.0002) | -0.0031*** (0.0002) | -0.0026*** (0.0002) | -0.0059*** (0.0001) | -0.0144*** (0.0002) | -0.0143*** (0.0002) |
| After 2000 |  |  | -0.1427*** (0.0006) | -0.1698*** (0.0006) | -0.1276*** (0.0005) | -0.0753*** (0.0010) | -0.0727*** (0.0017) |
| Average career age (0~50) |  |  |  |  |  | -0.0077*** (0.0001) | -0.0077*** (0.0001) |
| Interdisciplinary team |  |  |  |  |  | 0.0260*** (0.0008) | 0.0260*** (0.0008) |
| Weak-tie team |  |  |  |  |  | 0.0183*** (0.0008) | 0.0183*** (0.0008) |
| After2000*Remoteteam |  |  |  |  |  |  | -0.0032*** (0.0018) |
| Remote team+ After2000*Remote Team |  |  |  |  |  |  | -0.0088*** (0.0010) |
| Field controls |  |  |  | ✓ | ✓ | ✓ | ✓ |
| Author fixed effects |  |  |  |  | ✓ | ✓ | ✓ |
| Constant | 0.6874*** | 0.7124*** | 0.7969*** | 0.9065*** | 0.7872*** | 0.6491*** | 0.6465*** |
| N. of Observations | 2,903,964 | 2,903,964 | 2,903,964 | 2,903,964 | 9,031,126 | 3,821,308 | 3,821,308 |
| R-squared | 0.0044 | 0.0051 | 0.0236 | 0.0764 | 0.3404 | 0.3787 | 0.3787 |

**Extended Data Table 2. Assessing the robustness of declined disruption with increased collaboration distance in technology**. From our dataset of patenting teams, we selected 1,253,090 inventors who filed two or more patents. These inventors have filed 2,903,964 patents between 1976 and 2020, yielding 9,031,126 patent-author records. We use this dataset to build stepwise regression models in the same way as for scientific teams. We note that the remote work penalty—the negative relationship between remote collaboration and disruption—is robust across all specifications. When teams move from 0km to more than 600km collaboration distance, for example, the predicted disruption probability, holding other variables constant, declines from 58.2% to 56.5% (p-value < 0.001), or 2.9% in relative terms.

Note: All statistical tests are two-sided t-test and no adjustments were made for multiple comparisons. For Model 5-7, standard errors (in parentheses) are clustered at the inventor level. * p < 0.05; ** p-value < 0.01; *** p-value < 0.001. We used the REGHDFE package in STATA16[45] to implement the fixed-effects regressions.

|  | Probability to conceive | p-value | Probability to perform | p-value |
|---|---|---|---|---|
|  | Model 1 |  | Model 2 |  |
| Remote team | -0.0541*** <br> (0.0042) | <0.001 | 0.0159*** <br> (0.0037) | <0.001 |
| Team size (2~10) | -0.0358*** <br> (0.0011) | <0.001 | -0.0010 <br> (0.0009) | 0.306 |
| After 2010 | 0.0400*** <br> (0.0055) | <0.001 | -0.0456*** <br> (0.0049) | <0.001 |
| Field controls | ✓ |  | ✓ |  |
| Author fixed effects | ✓ |  | ✓ |  |
| Constant | 0.8261*** | <0.001 | 0.4139*** | <0.001 |
| N. of Observations | 65,143 |  | 65,143 |  |
| R-squared | 0.4995 |  | 0.5886 |  |

**Extended Data Table 3. Assessing the robustness of the reduced probability of conceiving research for the same scientist when switching from onsite to remote**. From our dataset of author contributions, we select 21,373 scientists who worked in both onsite and remote teams. These authors published 31,815 papers in total, which gives us 65,143 paper-author records. Using author-fixed effect regressions[40,45], we find that the same scientist is less likely to conceive research and more likely to perform experiments when switching from onsite to remote, when fields, periods, and team sizes are accounted for.

Note: All statistical tests are two-sided t-test and no adjustments were made for multiple comparisons. For Model 1-2, standard errors (in parentheses) are clustered at the author level. * $p < 0.05$; ** p-value $< 0.01$; *** p-value $< 0.001$. We used the REGHDFE package in STATA16[45] to implement the fixed-effects regressions.

|  | Probability to co-conceive | p-value | Probability to co-perform | p-value |
| --- | --- | --- | --- | --- |
|  | Model 1 |  | Model 2 |  |
| Remote team | -0.0196*** (0.0047) | <0.001 | 0.0151*** (0.0040) | <0.001 |
| Team size (2~10) | -0.0251*** (0.0017) | <0.001 | 0.0064*** (0.0014) | <0.001 |
| After 2010 | 0.0187* (0.0082) | 0.022 | -0.0470*** (0.0069) | <0.001 |
| Field controls | ✓ |  | ✓ |  |
| Author fixed effects | ✓ |  | ✓ |  |
| Constant | 0.5075*** | <0.001 | 0.2271*** | <0.001 |
| N. of Observations | 36,253 |  | 36,253 |  |
| R-squared | 0.6114 |  | 0.6803 |  |

**Extended Data Table 4. Assessing the robustness of the reduced probability of co-conceiving research for the same pair of scientists when switching from onsite to remote.** From our dataset of author contributions, we select 15,294 pairs of scientists who collaborate in both onsite and remote teams. These pairs of authors published 11,313 papers in total, which leaves us with 36,253 paper-author-pair records. Using fixed effect regressions to control the difference between author pairs[40,45], we confirm that the same pair of scientists are less likely to co-conceive research and more likely to co-perform experiments when switching from onsite to remote, when fields, periods, and team sizes are accounted for.

Note: All statistical tests are two-sided t-test and no adjustments were made for multiple comparisons. For Model 1-2, standard errors (in parentheses) are clustered at the author-pair level. * p < 0.05; ** p-value < 0.01; *** p-value < 0.001. We used the REGHDFE package in STATA16[45] to implement the fixed-effects regressions.